\documentclass[12pt]{article} \usepackage{epsfig} \usepackage{hhline}
\usepackage{amsmath,amssymb} \usepackage{boxedminipage}
\title{On the Worst-case Performance of the 
Sum-of-Squares Algorithm for Bin Packing}
\author{\sc Janos Csirik
        \thanks{
        {\tt  csirik@inf.u-szeged.hu}.
        Dept. of Computer Sciences,
        University of Szeged, Szeged, Hungary.
        }
\and
        \sc David S. Johnson
        \thanks{
        {\tt dsj@research.att.com}.
        AT\&T Labs - Research, Room C239, 180 Park Avenue,
        Florham Park, NJ 07932, USA.
	}
\and
	\sc Claire Kenyon
        \thanks{
        {\tt claire@cs.brown.edu}.
        Computer Science Department, Brown University,
        Providence, RI 02912.
        }
}

\begin{document}
\maketitle

\newtheorem{theorem}{Theorem} 
\newtheorem{freelemm}{Lemma}[section]
\newtheorem{lemma}{Lemma}
\newtheorem{coro}{Corollary}[theorem]
\newtheorem{obse}[theorem]{Observation}
\newtheorem{prop}[theorem]{Proposition} 
\newtheorem{claim}{Claim}[theorem]
\newtheorem{defi}{Definition}
\newtheorem{rema}[theorem]{Remark} 
\newtheorem{exam}[theorem]{Example}
\newtheorem{conj}[theorem]{Conjecture}
\renewcommand{\theequation}{\thesection.\arabic{equation}} 

\begin{abstract}
The Sum of Squares algorithm for bin packing was defined in~\cite{CJK99}
and studied in great detail in~\cite{sumsq2004}, where it was proved that its
worst case performance ratio is at most 3. In this note, we improve the
asymptotic worst case bound to $2.7777\ldots$
\end{abstract}

\section{Introduction}
%Bin packing def'n
In the classical bin packing problem, we are given an integer bin capacity $B$
and a list $L=(a_1,\ldots ,a_n)$ of items with each item $a$ having positive
size $s(a) \leq B$.
Our goal is to pack the items into
a minimum number of bins, i.e., partition them into a minimum number of
subsets such that the sum of the item sizes in each subset is $B$ or less.
This problem is NP-hard, so typically we must settle for approximation
algorithms, i,e., algorithms that produce packings with a relatively small
but possibly not minimum number of bins.
Of special interest are on-line algorithms, i.e., ones that assign each item
in turn to a bin without knowledge of the sizes or number of future items.

%SS def'n
The {\em Sum of Squares} on line bin packing algorithm (SS), introduced in
\cite{CJK99}, is applicable to instances where the item sizes are integral, and
is surprisingly effective whenever the item sizes are independent
identically distributed random variables~\cite{sumsq2004}.
It uses the following simple rule to add an item to the current packing $P$.
Let $s$ be the size of the item and
let $ss(P)=\sum_1^{B-1}n_h(P)^2$, where $n_h(P)$ is the number of bins in
$P$ whose {\em level} (the total size of the items the bin contains)
is equal to $h$.
Then the item is placed into either a new bin or a partially filled
bin with level less that or equal to $B-s$, with the choice made so as
to minimize $ss(P')$ for the resulting packing $P'$.
In what follows, we will write $n_h$ instead
of $n_h(P)$ when the packing under study is clear from the context.

%Known det. lower bounds (1.54) and upper bounds (3)
For any list $L$, let $s(L)=\sum_i s(a_i)$.
Clearly, the number of bins must be at least $\lceil s(L)/B \rceil$.
In~\cite{sumsq2004}, it was proved that the number $SS(L)$ of bins
used by SS is at most $3\lceil s(L)/B\rceil$,
and hence SS has an asymptotic performance ratio of at most 3.
In addition, instances were presented that implied that the
asymptotic performance ratio for SS is at least 2.
%Our main theorem
In this note, we give an improved worst-case analysis of SS 
that begins to close the gap, lowering the asymptotic worst-case
performance ratio from 3 to $25/9 = 2.7777\ldots$ 
\begin{theorem}\label{thm:wc}
For all lists $L$,
$$
SS(L) ~<~ \frac{25}{9} \cdot \frac{s(L)}{B} + 2 ~\leq~
\frac{25}{9} OPT(L) + 2.
$$
\end{theorem}

%Discussion
{\bf Discussion.}
We expect that this new bound can be further improved.
The proof of the original factor-of-3 bound in~\cite{sumsq2004}
was based on examining the last
time {\em one} item of a certain type was inserted, whereas our analysis
here is based on examining the last time {\em two} particular items were
inserted.
Extending this analysis to three or more items might yield further 
improvements, although we expect that the tradeoff between
bound-improvement and the length of the proof will follow
the law of diminishing returns.
Major improvements will probably require significant new ideas.

\section{Proof of Theorem~\ref{thm:wc}}

The proof repeatedly uses the following straightforward key property
of the Sum of Squares algorithm, which was already used in the
proof of the factor-of-3 result in~\cite{sumsq2004}.

\begin{lemma}{\rm\label{lemma:sequence}\cite{sumsq2004}}
If SS starts a new bin when adding an item of size $s$ to the current packing
$P$, then, for any $j$, $1 \leq j < B-s$, we have
$$n_{j}(P)\leq n_{j+s}(P).$$ 
\end{lemma}

Let $0<\alpha < \delta \leq 1/2$ be two parameters satisfying
the following inequalities.
\begin{equation} \label{ineq1}
2\alpha \leq 1-2\delta ,
\end{equation}
\begin{equation}\label{ineq3}
\frac{1+\alpha}{3} \geq \delta , \hbox{ and}
\end{equation}
\begin{equation}\label{ineq4}
\alpha \leq 2 -\frac{16}{3}\delta.
\end{equation}
We will show that, with the possible exception of two bins,
the bins of the SS packing are filled to an average level $\delta B$.
This implies $s(L)/B \geq \delta (SS(L)-2)$.
Maximizing $\delta$ under the above constraints yields
$\delta = 9/25$ (with $\alpha = 2/25$), hence the theorem.
Since our proof relies only on the fact that $\alpha$ and $\delta$
satisfy (\ref{ineq1}) through (\ref{ineq4}), we can also conclude
that no better bound can be obtained using the same basic proof technique.

\smallskip
If a bin was ever started with an item of size less than
 $\delta B$, let the last item to start
such a bin be $x$ and let its size be $s < \delta B$.
In addition, if a bin was ever started with an item of size less than
or equal to $\alpha B$, let the last item to start
such a bin be $x'$ and let its size be $s' \leq \alpha B$.

If $x'$ exists, for each $j\in [1,s']$, we define $c_j$ to be the largest integer
such that $j+c_js' < B$.
Notice that $j+c_js'>B(1-\alpha)$.
We now break the proof into cases.

\smallskip
{\bf Case 1: $x$ does not exist.} Then all non-empty bins contain at
least one item of size $\geq \delta B$ and we are done.

\smallskip
{\bf Case 2: $s\leq \alpha B$.}
Then $x'=x$ is also the last item of size
$\delta B$ or less to start a new bin.
We analyze the packing at the time when $x=x'$ was packed.
Call a level {\em small} if it is less than $\delta B$ and {\em large}
if it exceeds $\delta B$.
We pair every small level with a large level 
which is in the same congruence class mod $s'$:
level $j +is'$ is paired with level $j+(c_j-i)s'$, for
$1 \leq j \leq s'$ and $j + is' < \delta B$.
To confirm that the second level is indeed large, note that
\begin{equation}
\frac{(j+is') + (j+(c_j-i) s')}{2}
= \frac{2j+c_js'}{2} \geq \frac{B(1-\alpha)}{2}\geq \delta B,
\end{equation}
where the last inequality follows from~(\ref{ineq1}).
Thus if the first level is small, the second is
indeed large.
Moreover, any combination of a bin of the first level with
a bin of the second will have average contents at least $\delta B$.

Moreover,  Lemma~\ref{lemma:sequence} 
implies that $n_{j+(c_j-i)s'} \geq n_{j+is'}$,
and so we can assign each bin with level $j+is'$ 
as the {\em unique}  mate of a bin with level $j+(c_j-i)s'$.
It follows that at the time when $x'$ was packed we had
$s(L)/B \geq \delta\cdot  SS(L)$.
The new bin into which $x'$ was packed may be less full,
but thereafter no bin can decrease in level and all subsequent new bins
must have level at least $\delta B$, so in the end we must have
$s(L)/B > \delta (SS(L)-1)$, which implies the theorem.

\smallskip
{\bf Case 3: $s > \alpha B$.}
Then $x'\neq x$ or $x'$ does not exist.
We again exploit the concept of { mate}.
If $x'$ exists, let $m\geq 1$ be the largest integer
such that $ms' \leq \alpha B$.
Let $\Delta = ms' \geq \alpha B/2$.
Note that $ms' < \delta B$ and so using the same scheme we used
in Case 2, we can pair each bin with level $\leq ms'$ with a
bin with level $\geq \delta B$ for which it is the unique mate.
(In what follows, we shall refer to each bin in the pair as the
mate of the other.)
Moreover, here we have a slightly stronger result:
at the time when $x'$ was packed,
we had for each $j$, $1 \leq j \leq s'$, that
every bin with level $j+is'\leq ms'$
had as its mate a bin whose contents had total size at least $B(1-\alpha)$.
To see this, note that since $j+is'\leq ms'$ and $j>0$, we have $i\leq m-1$,
and so:
\begin{equation}\label{matebound}
j+(c_j-i)s'\geq j+c_js-(m-1)s'\geq B-s'-(m-1)s'\geq B(1-\alpha).
\end{equation}
If $x'$ does not exist, let $\Delta = \alpha B > \alpha B/2$
and note that at the time $x$ was packed we
had $n_h = 0$ for all $h \leq \alpha B$,
and hence there are no mates. 

We now analyze the packing at the time when $x$ is packed,
given that at that time all bins with levels $\leq \Delta$ were
mates of bins at levels that satisfied (\ref{matebound}) and
that $\Delta \leq \alpha B$.
We partition the bins with levels $>\Delta$ into congruence classes mod $s$:
For each $h\in [\Delta +1,\Delta +s]$, class $D_h$ consists of the bins
with levels $h$, $h+s$, $h+2s$, $\dots$, $h+d_hs$, where $d_h$ is the
largest integer such that $h+d_hs< B$.
Notice that $d_h \geq 1$ follows from~(\ref{ineq1}) and the
fact that $s < \delta B$.
For each class $D_h$, we add as {\em honorary} members those bins
with levels $\leq \Delta$ that are mates of bins in $D_h$.
Thus at the time $x$ is to be packed,
every bin is either in a class $D_h$ or an honorary member of such a class,
with the possible exception of the bin that received $x'$.

We shall now show that for each $h$, the bins in $D_h$,
together with the honorary members of $D_h$, have average
content at least $\delta B$.
After the bin containing $x$ is started, all subsequent bins will start
with items of size $\delta B$ or greater, so we
will thus be able to conclude
that $s(L)/B > \delta(SS(L)-2)$ and the theorem will follow.

{\bf Subcase 3.1.}
Let us first consider the case when $D_h$ contains
no honorary members.
By Lemma~\ref{lemma:sequence}, 
the average content for bins in $D_h$ is at least
\begin{equation*}
\frac{1}{d_h+1}\sum_{i=0}^{d_h}(h +is) = h + \frac{d_hs}{2}.
\end{equation*}
Our analysis will now use the bounds 
$h+(d_h+1)s\geq B$, $s<\delta B$, and $h>\Delta >\alpha B/2$,
which follow from the definitions of $d_h$, $s$, and $\Delta$.
We also use the 
facts that $d_h$ is an integer and that, as a consequence of (\ref{ineq1})
and (\ref{ineq3}), we have $\delta \leq 3/8 < 2/5$.

If $d_h=1$, then
$$h+s/2 > (B-2s)+s/2=B-3s/2 > B-3\delta B/2 > \delta B.$$  

If $d_h=2$, then we use assumption~(\ref{ineq3}) to get
$$h+s > h+(B-h)/3=(B+2h)/3 > (B+\alpha B)/3 > \delta B.$$

Finally, if $d_h\geq 3$, then
$$h+\frac{d_hs}{2} > h+\frac{d_h}{ 2}\left(\frac{B-h}{d_h+1}\right) >
\frac{d_hB}{2(d_h+1)}\geq \frac{3}{8}B\geq \delta B.$$

{\bf Subcase 3.2.}
Let us now consider the case when $D_h$ does
contain honorary members.
By (\ref{matebound}) the bins of $D_h$ which have mates
have levels greater than or 
equal to $B(1-\alpha)$, and so clearly we must have $h+d_hs \geq B(1-\alpha)$.
Since $s> \alpha B$, only this last level $h+d_hs$ in $D_h$ 
can contain bins that have mates.
Thus there are at most $n_{h+d_hs}$ honorary members.
Moreover, we claim that $d_h\geq 2$, since otherwise
we would have $h+s\geq B(1-\alpha )$ 
and hence 
$$B(1-\alpha )\leq h+s \leq \Delta +2s < \alpha B + 2\delta B,$$
in contradiction with~(\ref{ineq1}).

Let us partition the true and honorary members of
$D_h$ into subclasses $E_i$, $0 \leq i \leq d_h$, as follows:
$E_0$ consists of the first $n_h$ bins in each level of $D_h$, together
with the first $n_h$ honorary members of $D_h$ (or all such
bins if there are fewer than $n_h$ of them).
Inductively, for each $i >0$,
$E_i$ consists of the first $t_i = n_{h+is} - n_{h+(i-1)s}$
as-yet-unassigned bins in each level, a non-negative number
by Lemma~\ref{lemma:sequence},
together with the first $t_i$ unassigned honorary members of $D_h$
(or all such bins if there are fewer than $t_i$ of them).
Note that  $E_i$  contains 
$t_i$ bins for each level $h+ks$, $i \leq k \leq d_h$,
plus up to $t_i$ honorary bins.
Thus $|E_i| \leq t_i(d_h-i+2)$.

Let $X_i=\sum_{k=i}^{d_h}(h+ks)/(d_h-i+2)$.
This is the average level of a collection of bins,
one empty and one having level $h+ks$, $i\leq k \leq d_h$.
It is easy to see that $X_i$ is a lower bound on the average contents
of the bins in $E_i$ when $E_i$ is nonempty.
We will prove that for every $i$, $X_i \geq \delta B$.
This will imply that the average contents for all the bins in $D_h$
is at least $\delta B$.
Since this will be true for all $h$, the theorem will follow.

First observe that as long as $h+is \leq \delta B$, we have
that $X_{i}\geq \delta B$ implies $X_{i+1}\geq \delta B$ 
since $X_{i+1}$ can be obtained from $X_{i}$ by removing
a bin of level $h+is \leq \delta B$.
Similarly, as long as $h+is>\delta B$, we have that
$X_{i+1}\geq \delta B$ implies $X_i \geq \delta B$
since $X_{i}$ can be obtained from $X_{i+1}$ by adding
a bin of level $h+is>\delta B$.
Hence it is enough to prove the claim for $i=0$ and for $i=d_h$.

If $i=d_h$, we have $X_i=(h+d_hs)/2\geq B(1-\alpha )/2 \geq  \delta B$
by (\ref{ineq1}), and we are done.

If $i=0$ we have:
$$X_i\geq\frac{1}{d_h+2}\sum_{k=0}^{d_h}(h+ks)
=\frac{d_h+1}{d_h+2}\left(h + \frac{d_hs}{2}\right)
\geq \frac{d_h+1}{d_h+2}\left(h + \frac{(B(1-\alpha)-h)}{2}\right)$$
by  (\ref{matebound}). Using
 $h>\Delta \geq \alpha B/2$, $d_h\geq 2$, and assumption (\ref{ineq4}) gives
$$X \geq \frac{d_h+1}{d_h+2}\left(\frac{B(1-\alpha/2)}{2}\right)
\geq \frac{3}{8} (1-\alpha /2) B
\geq \delta B $$
and the theorem is proved.  $\square$

\bibliographystyle{plain}
\bibliography{bp}
\end{document}